\documentclass[referee]{aa}

\usepackage{graphicx,epsf,psfig,epsfig,times,amsmath,amssymb}

\usepackage{natbib}
\bibpunct{(}{)}{;}{a}{}{,}

\def \lowhard {{\it low/hard}}
\def \hardpl {{\it hard/PL}}
\def \intermediate {{\it intermediate}}
\def \soft{{\it soft}} 

\def\nh {N_{\rm H}}

\def\comptt{{\sc Comptt}}
\def\comptb{{\sc Comptb}}
\def\bmc{{\sc bmc}}
\def\17p2{\mbox{GX~17+2}}

\def\gaussian{{Gaussian}}
\def\chiq{$\chi^2$}

\def\cm2{cm$^{-2}$}
\def\s1{s$^{-1}$}

\def\wabs{{\sc wabs}}
\def\sax{{\it BeppoSAX}}
\def\xte{{\it RXTE}}
\def\integral{{\it INTEGRAL}}

\def\xspec{{\rm  XSPEC}}

\def\cygx2{\mbox{Cyg~X--2}}
\def\scox1{\mbox{Sco~X--1}}

\def\kts{kT_{\rm s}}
\def\kte{kT_{\rm e}}
\def\rbb{R_{\rm bb}}

\def\comptt{{\sc comptt}}

\def\chiq{$\chi^2$}

\begin{document}

\title{The X-ray spectral evolution of Cyg X--2 in the framework of bulk Comptonization}

\author{R. Farinelli\inst{1}, A. Paizis\inst{2}, R. Landi\inst{3} \& L. Titarchuk\inst{1, 4}}
\offprints{R. Farinelli, farinelli@fe.infn.it}

\institute{
Dipartimento di Fisica, Universit\`{a} di Ferrara, Via Saragat 1, 44100 Ferrara, Italy
\and INAF-IASF, Sezione di Milano, Via Bassini 15, 20133 Milano, Italy
\and INAF-IASF, Sezione di Bologna, Via Gobetti 1, 40100, Bologna, Italy
\and NASA/GSFC, Greenbelt, MD 20771}

  \abstract {The interplay of thermal and bulk motion Comptonization to 
explain the spectral evolution of neutron star LMXBs including transient hard X-ray 
tails is gaining a strong theoretical and observational support.  
The last momentum has been given by the advent of a new 
XSPEC Comptonization model, \comptb, which  includes 
 thermal and bulk Comptonization.}
   {We used \comptb\ to investigate the spectral evolution of the neutron star LMXB \cygx2\ along 
its Z-track. We selected a single source in order to trace in a quantitative way
the evolution of the physical parameters of the model. }
   { We analyzed archival broad-band \sax\ spectra of \cygx2. Five broad-band spectra have been 
   newly extracted according to the source position in the Z-track described in the color-color 
   and color-intensity diagrams.}
 { We have fitted the spectra of the source with two \comptb\ components. The first one, with bulk 
 parameter $\delta$=0, dominates the overall source broad-band spectrum and its origin  is  related to  thermal upscattering (Comptonization)
 of cold seed photons off warm electrons in high-opacity  enviroment.
 We attribute the origin of these seed  photons to the part of the disk which illuminates  the outer
 coronal region (transition layer) located between the accretion disk itself and the neutron star surface.  This thermal component is roughly constant with time and with inferred mass accretion rate. The second \comptb\ model describes the overall Comptonization (thermal plus bulk,  $\delta >$ 0) of hotter seed photons  which come  from both  the inner transition layer and from the neutron star surface. 
The appearance of this component
in the colour-colour or hardness-intensity diagram is  more pronounced  in the horizontal branch and is progressively disappearing towards the normal branch, where a pure blackbody spectrum is observed.}
{The spectral evolution of \cygx2\ is studied and interpreted in terms of changes in the 
innermost environmental conditions of the system, leading to a variable  thermal-bulk 
Comptonization efficiency.}
   

\authorrunning{Farinelli et al. }
\titlerunning{X--ray spectral evolution of Cyg~X--2}
\maketitle


\section{Introduction}
\label{introduction}
The  bright persistent Low Mass X-ray Binary (LMXB) \mbox{Cyg X--2} is one of the Galactic sources hosting a neutron star (NS) which belong to the Z class, due to the   Z-shape it displays in its color-color (CD) and hardness-intensity diagram
(HID). The parts of the diagram  
are referred to as the horizontal branch (HB) at the top of the Z, 
the normal branch (NB) along the diagonal, and the flaring branch (FB)
at the base of the Z diagram.  
Actually, Z sources move continuously along this diagram and the position 
of the source within the Z is believed to be related to the mass accretion rate that
increases from the HB, lowest accretion, to the FB, highest accretion 
\citep{hasinger90}.

\begin{figure*}
\centering
\includegraphics[width=0.49\linewidth]{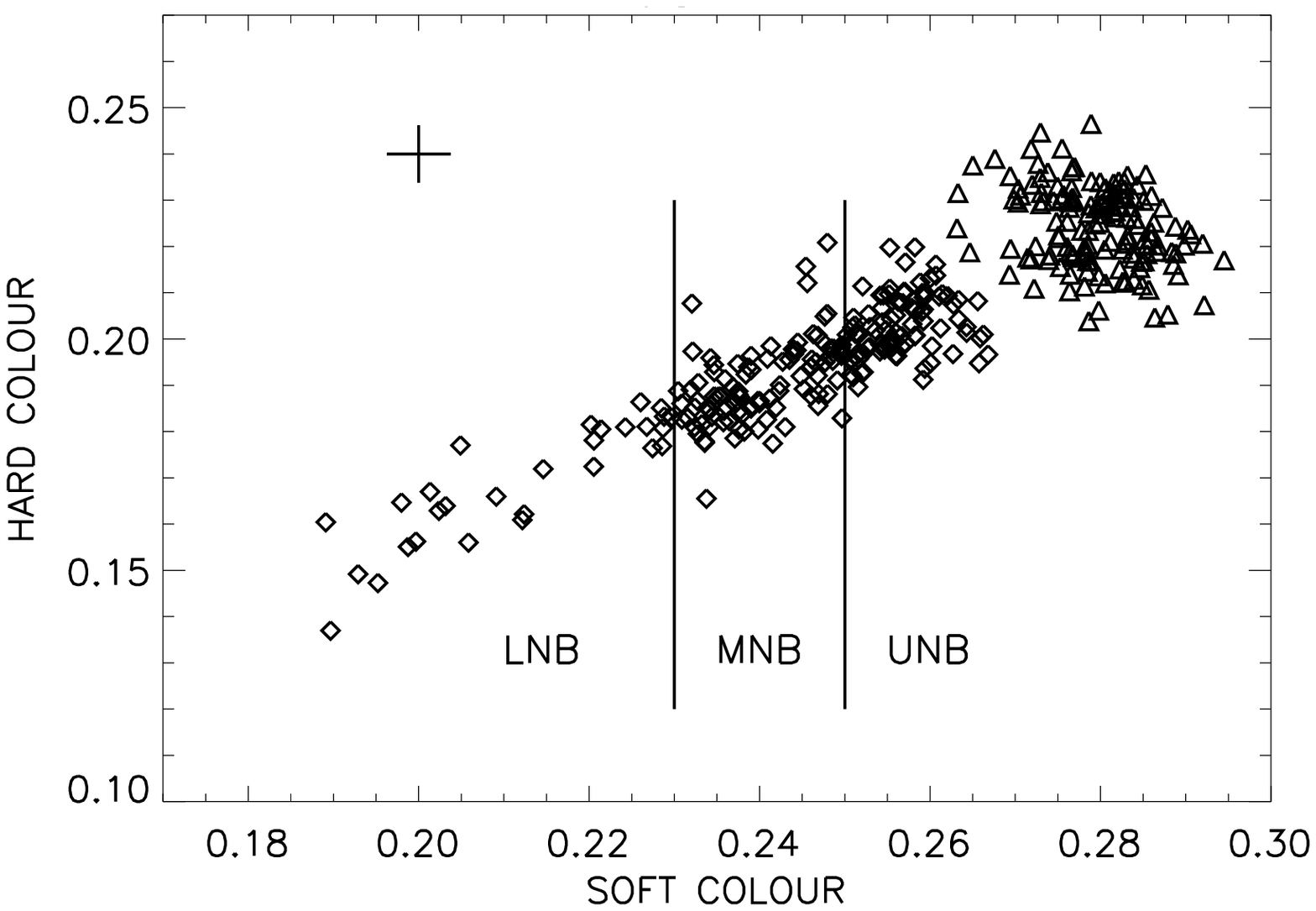}
\includegraphics[width=0.49\linewidth]{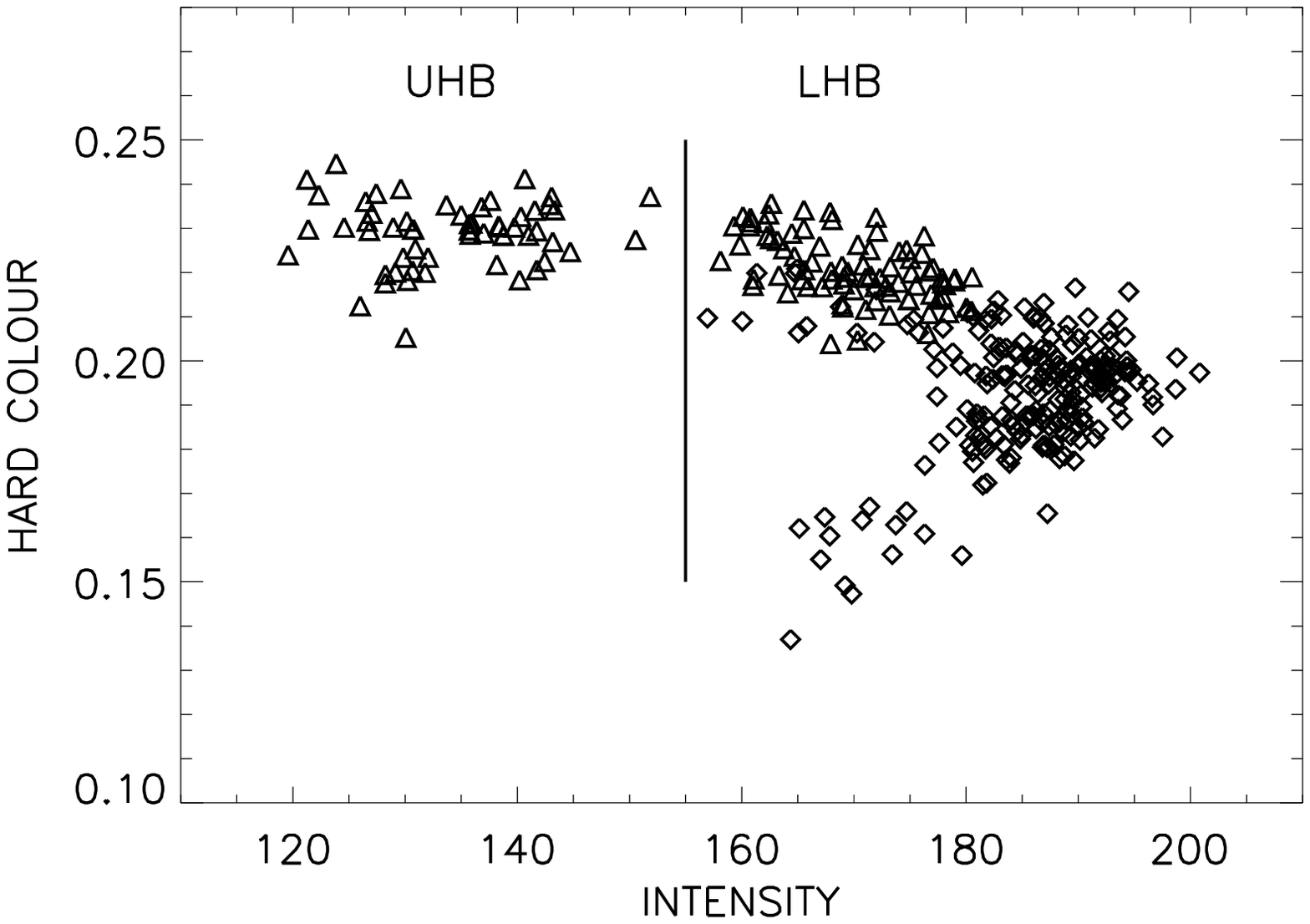}
\caption{Color-color diagram ({\it left panel}) and hardness intensity diagram ({\it right panel}) of Cyg~X--2 during the 1996 ({\it triangles}) and 1997 observation ({\it diamonds}).  
The hard colour is defined as the \mbox{7--10.5 keV/4.5--7 keV} bands count rate ratio while the 
soft colour is defined as the \mbox{4.5--7 keV/1.4--4.5 keV} count rate ratio.  For the hardness intensity diagram, 
the intensity is given by the \mbox{1.4--10.5 keV} count rate.  A typical error bar is shown in the left panel. The regions of the branches for which particular spectral analysis was carried-out are also shown. 
\label{hid}}
\end{figure*}

The $<$20\,keV X-ray spectrum of Cyg~X--2, and of persistent bright NS LMXBs in general,
has been usually  described 
as the sum of a soft and a hard component. The soft component is interpreted as 
emission from the accretion disc \citep[Eastern model, ][]{mitsuda84}
or originating at (or close to) the NS \citep[Western model, ][]{white86}
whereas the hard component is most likely formed  due to Comptonization of NS and/or disc emission 
by a hot plasma (so-called "corona") electrons. These two models describe equally well the spectra of NS LMXBs below about
20\,keV. The first studies of the \mbox{Cyg X--2} X-ray spectrum above $\sim20$\,keV were performed with
detectors on balloons \citep{peterson73}.
Interestingly, an unexpected hardening was observed in the spectrum of \cygx2\ that
was fitted by a powerlaw (PL) with photon index of 2.8 \citep{maurer82} or 1.9 \citep{ling96}.
It soon became clear that simultaneous broad band observations of
the X-ray spectrum were needed to investigate the nature of this spectral flattening. \\
The advent of broad-band X-ray missions, such as \textit{BeppoSAX, RXTE, INTEGRAL},
revealed that many such spectral hardenings (so-called "hard tails") occur in 
 Z sources: Cyg~X--2 \citep{frontera98,disalvo02b}, GX~17$+$2 \citep{farinelli05,disalvo00}, 
GX~349$+$2 \citep{disalvo01}, Sco~X--1 \citep{damico01,disalvo06}, GX~5--1 \citep{paizis05,asai94}, 
GX~340$+$0 \citep{lavagetto04}. Recently, a hard tail has also been discovered in 
the bright atoll source GX~13$+1$ \citep[][hereafter P06]{p06}.\\
The spectra of these sources and in particular the hard X-ray tails
 have been extensively studied and focused on a single source basis,   and 
mainly in terms of phenomenological models  \citep[see][for a review on NS LMXB spectra]{barret01, disalvo02}. In the attempt to 
study these sources in terms of a \emph{unified physical scenario} in the less known domain above
20\,keV, P06 studied the long term average hard X-ray ($>$20\,keV) 
   spectra of a sample of twelve bright NS LMXBs (six Z and six atoll sources), 
   using data from IBIS instrument
   on-board  
   \textit{INTEGRAL} \citep{winkler03}. 
Merging their results with those of \cite{falanga06} for the atoll source 
\mbox{4U~1728--34} (GX~354--0), P06
identified four main spectral states for NS LMXBs (see Fig. 4 in P06): \lowhard\ 
state (\mbox{GX~354--0}), \hardpl\ state  (\mbox{H~1750--440} and \mbox{H 1608--55}), 
\intermediate\ state (where the hard X-ray tail appears, e.g., Cyg~X--2, Sco~X--1, \mbox{GX~5--1}, GX~17$+$2) and 
 \soft\ state (e.g., GX~3$+$1, GX~9$+$1, GX~9$+$9). 
The different spectral states, including the hard tails, could be well fit in 
terms of the interplay of 
thermal and bulk Comptonization (TC and BC, respectively) using the \bmc\ model in \xspec. 
The relative contribution between the two Comptonization regimes (thermal versus bulk)
is proposed to be drawn by the \emph{local} accretion rate $\dot{M}$. Indeed (see P06), starting 
from the 
lowest level $\dot{M}$, the \lowhard\ state spectra, whose  cut-off is  below
100\,keV, can be interpreted in terms of  TC of soft photons off a hot ($\sim$30\,keV) electron population; 
at increasing $\dot{M}$, bulk Comptonization starts to become relevant 
with the result of moving at higher energies the cut-off (\hardpl\ state); higher accretion
rates lead  to high bulk inflow Comptonization efficiency
 that is seen as  an  extended hard tail in the spectrum above 30\,keV (\intermediate\ state). 
In this state TC becomes less efficient since 
the coronal plasma is cooled 
    down to $\sim$3\,keV and the emergent spectrum is dominated by a strong thermal bump (with cut-off energy around 10 keV) with overlapped a high-energy PL-like tail up to 100 keV; finally, the 
\soft\ state spectra, with highest $\dot{M}$, are described by a single TC component (the ``bump'') with a low-$\kte\ $ and  high-$\tau$ plasma, as expected
in a high $\dot{M}$ environment. The emission above 
30\,keV is no longer present since the high local pressure gradient either prevents matter  to reach 
the NS or strongly decelerates it\footnote{See P06 for a more detailed description of the X-ray spectral evolution and for the X-ray / radio correlation obtained.}.\\
The study of the average spectra of the twelve NS LMXBs above 20\,keV allowed P06 to present a 
qualitative scenario for the X-ray spectral evolution of these sources along with 
the radio - X--ray connection. Nevertheless, the lack of data below 20\,keV  
 prevented the authors from drawing more stringent conclusions on the 
 accretion geometry of the systems and on the physical parameters.
To overcome these limitations and to 
obtain a quantitative view of the parameter evolution within a single source
we present the study of the broad-band 
(0.4-120\,keV) \textit{BeppoSAX} 
spectra of Cyg~X--2 in terms of a newly updated version of the \bmc\ model, hereafter 
referred to as \comptb\ \citep{farinelli08}.
A similar analysis of \sax, \xte\ and \integral\ data from a number of LMXBs  has been already  made by
\cite[][hereafter F07 and F08, respectively]{farinelli07,farinelli08}. 

\section{Observations and data analysis}

\begin{figure}
\centering
\includegraphics[width=1.0\linewidth]{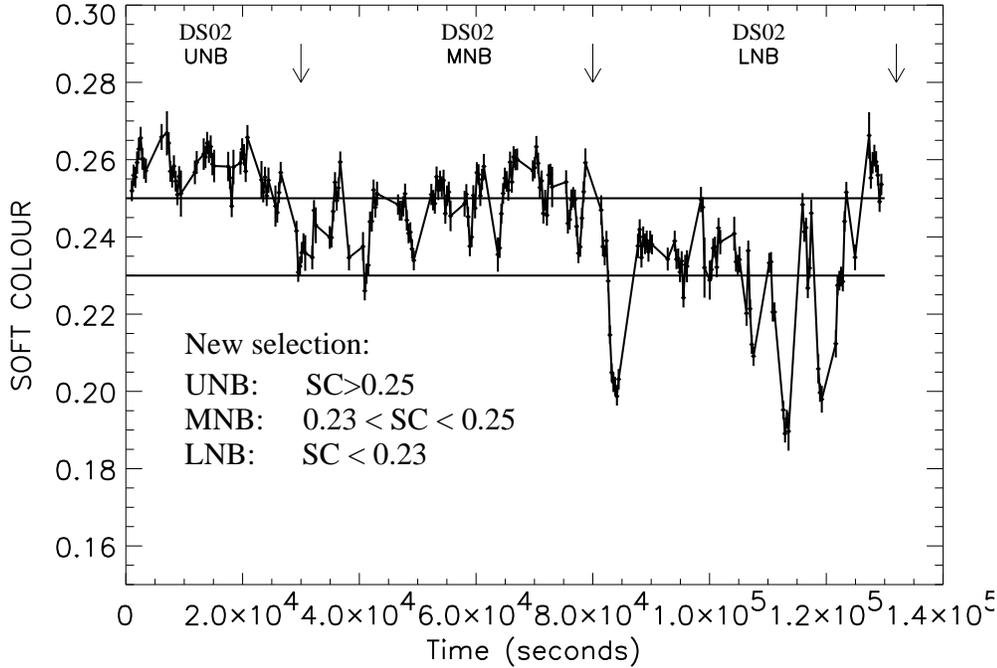}
\caption{Temporal evolution of Cyg~X--2 soft color in the NB.  The three  intervals 
used by DS02 to classify the NB  are shown with the upper arrows. The soft color intervals used in this work are given and shown in the horizontal lines.
\label{cvstime}}
\end{figure}

We analyzed two \textit{BeppoSAX} \citep{boella97a} observations of Cyg~X--2, the first one
performed on 1996 July 23 (00:54:21 to 23:53:06 UT) and the second on 
1997 October 26 (15:37:11 UT) to October 28 (04:00:12 UT). This  data set has already been presented 
by Di Salvo et al. (2002, hereafter DS02). 
In the analysis carried-out by DS02, colour-colour  and hardness intensity diagrams (CD and HID, respectively) of the source with the {\it Medium-Energy Concentrator Spectrometer} \citep[1.8-10 keV;][]{boella97b}.
were produced. The soft colour (SC) was defined as the 4.5--7\,keV/1.4--4.5\,keV count rate ratio,
while the hard colour  by the 7--10.5\,keV/4.5--7\,keV count rate ratio. On the other hand, in the HID 
the  intensity was defined by the 1.4--10.5 keV count rate.
During the 1996 observation, the source was at HB of the CD/HID diagram; this is much more evident using  the HID, were two distinct regions separated by a gap around 150 cts \s1 are clearly visible (see Fig.~\ref{hid} here, right hand panel, and Fig.~1 in DS02). These two HB regions  were labeled as an upper HB and a lower HB
(UHB and LHB, respectively) and two broad-band spectra were separately extracted for each of them. 
We adopted the same approach.
On the other hand, during the 1997 observation, the source was at NB; in this latter case, the NB structure
is more evident using the CD. Three spectra were extracted by DS02 for the NB, and they were labeled as an upper NB, a medium NB and a lower LNB (UNB, MNB and LNB, respectively, see Fig.~2, left panel, in DS02).
We note however that the time-filters used by DS02 to extract the source spectra of the three parts of the NB were
almost rough as they actually simply considered three consecutive time intervals which were associated
to the periods spent by the source \emph{on average} at UNB, MNB and LNB stages, respectively.
We realize however that this is, in fact, not true as the source motion along the NB is not strictly continuous and smooth but it actually behaves like a random motion; this is clearly evident plotting e.g. the SC as a function of time (see Fig.~\ref{cvstime}).
In order to produce spectra which correspond to \emph{effective} UNB, MNB and LNB positions we thus used a more refined criterion; namely, we also divided the NB in three intervals (see Fig.~\ref{hid}, left panel), according to SC-value being greater than 0.25 (UNB), between 0.23 and 0.25 (MNB) and lower than 0.23 (LNB), but temporal filter for spectral extraction were produced  following the source behaviour as reported in Fig.~\ref{cvstime}.\\
The shift of the source position in the CD/HID between our plots and those reported in DS02 has two reasons: first, we extracted light-curves (such as energy spectra) from a 4$\arcmin$ region centered around the MECS source image (while in DS02 no spatial selection was used for producing CD/HID), and second we used a  better MECS instrumental channel-to-energy conversion law, required when extracting  light curves in different energy bands.

The time-filtered spectra of  the high-energy {\t Phowswich Detection System}  \citep[PDS;][]{frontera97} were produced using the \emph{XAS} package and we grouped the PDS channels in order to have S/N $\ga$ 3. Bin with lower threshold, as being statistically meaningless, were discarded and not included in the fit.
A further comment is required for the use of the LECS response files (the energy redistribution matrix and the effective area files, RMF and ARF, respectively). 
When the LECS count rate is about 50-60 cts \s1 or lower, the standard on-line available response matrix can be safely used, while for higher count rates
specific observation-related RMF and ARF files should be produced.
The fits on the LECS Crab spectrum ($\sim$ 200 cts \s1) clearly shows the presence of $\la$ 5\% residuals in the region 
0.5-1 keV and 2-3 keV\footnote{See http://bepposax.gsfc.nasa.gov/bepposax/software/index.html.}. 
It is thus evident that when a source spectrum is very bright and far from being Crab-like, these instrumental features can be significantly 
enhanced. This problem may be critical for LMXBs  as their low-energy X-ray emission is dominated by the presence of the soft $\sim$ 0.5 keV BB-like component in 
addition to photoelectric absorption.
The LECS count rate of the 1996 observation (HB spectra) of \mbox{Cyg X--2} was around 50 cts \s1 ($\sim$ 0.25 Crab)  while during 1997 observation (NB spectra) it
was on average $\sim 90$ cts \s1 ($\sim$ 0.4 Crab). We thus  produced new RMF and ARF files for each of the five selected spectra using the \emph{LEMAT} package. 
A fit on the Crab LECS spectrum shows that 0.8\% of systematic plus a fake 2.2 keV Gaussian emission line provide 
a reduced \chiq\ equal to 1. The fake excess around 2.2 keV is, in fact, observed not only in LECS but also in MECS spectra, 
and is the result of a strong edge-like decrease in the instrument effective area for both instruments, which becomes evident 
at very high instrument count rates; this excess is  \emph{observed in all our five spectra}. While  keeping in mind the 
 instrumental issues mentioned above, we added a  0.5\% systematic to all instruments; this is a good compromise between taking into account calibration uncertainties
 and avoiding unreasonably low reduced \chiq\ values in the best-fit models. 
Moreover, we added a Gaussian  emission line in the critical edge region ($\sim$ 2.2 keV) of LECS and MECS.
Another issue at low energies concerns the strong feature observed in the Cyg~X--2 LECS spectra (especially in the 1997 data-set) around 1\,keV.
Since its first detection during a rocket flight in 1971 \citep{bleeker72}, 
several experiments have reported this emission. 
While from the {\it Ariel V} data \citep{br84} it could not be possible 
to estimate the line equivalent width (EW) because of the low instrumental sensitivity, 
{\it Einstein} observations \citep{vrtilek88} suggested that this feature is likely due to a 
combination of unresolved Fe L-shell line emission that they identified with Fe XVII, 
O VIII/Fe XVII, Fe XX/Ni XX and Fe XVII/Fe XXII-XXIV features ranging from 0.74 to 
1.12 keV. Also {\it EXOSAT} observations supported this 
hypothesis \citep{chiappetti90}, although the data did not allow to resolve this 
complex region.  Based on subsequent measurements with the Broad Band X-ray Telescope 
(BBXRT) and {\it ASCA}, Smale et al. (1993,1994) modeled the excess with 
a broad Gaussian line having a full width half-maximum (FWHM) and EW in the range $\sim$ 
0.05-0.33 keV and 8-60 eV, respectively. This modeling provides a good 
description of the excess also in \sax\ observations \citep{kuulkers97}, resulting in an energy centroid of 1.02 keV, a $\sigma$ $\sim$ 0.47 and an EW 
of 74 eV. By assuming the model proposed by \cite{vrtilek88}, Kuulkers 
and co-workers found similar results but differences in EW, thus confirming that the 
shape of the feature, as well as its EW, changes from observation to observation.
However it is worth pointing out that both {\it Chandra} \citep{takei02} and {\it XMM-Newton} \citep{costantini05} observations of \mbox{Cyg X--2} revealed a complex structure of absorption 
edges (due to combination of N, O, Fe-L, Ne-K shells) in the region 0.4-0.8 keV but no emission around 1 keV.
Following the high energy-resolution of these two satellites at low energy and keeping in mind that sometimes the presence of a strong absorption edge, if not taken into account, may give rise of emission-like features in the residuals, we tried to use a single absorption edge in our best-fit models, but the result 
was completely unsatisfactory requiring actually a Gaussian emission line.
Investigating the origin (physical or instrumental) of such 1 keV emission feature is beyond the scope of the present
paper and all of the results  reported include a Gaussian emission line.
Independently from the physical or instrumental origin of the 1 keV emission line, we note that its presences 
strongly influences the determination of both the interstellar absorption $\nh$ and the temperature of the soft BB-like component. In particular, 
we find that too low $\nh$-values are 
obtained for the latter parameter ($\sim 0.08 \times 10^{22}$ cm$^{-2}$) if compared to  results of radio map measurement \citep[$\sim 0.19 \times 10^{22}$ cm$^{-2}$:][]{dl90},  XMM observations \citep[$\sim 0.18 \times 10^{22}$  cm$^{-2}$;][]{costantini05}, or \sax\ observation with no line included ($\sim 0.25 \times 10^{22}$ cm$^{-2}$; DS02, F08).
The difference of $\nh$ are more pronounced in the NB spectra with respect to HB ones:  we thus fixed $\nh= 0.15 \times 10^{22}$ cm$^{-2}$  for all five analyzed spectra, in order to keep as low as possible this model-dependent variations.

\begin{table*}[]
\begin{center}
 \caption[]{Best-fit parameters of the multi-component model \wabs(\comptb\ + \comptb\ + \gaussian).  Errors are computed at 90\% confidence level for a single parameter. Additional emission lines at $\sim$ 1 keV and $\sim$ 2.5 keV were included in all the five spectra (see Section 2).}
 \begin{tabular}{cccccc}
\hline
\hline
\noalign{\smallskip}
Parameter   &  UHB  &	LHB  &    UNB  &   MNB  & LNB \\
\noalign{\smallskip}
\hline
\noalign{\smallskip}
$N_{\rm H}^{\rm a}$ &[0.15]   &[0.15]   & [0.15]   &  [0.15] &  [0.15] \\
\noalign{\smallskip}
\hline
\noalign{\smallskip}
\multicolumn{6}{c}{\comptb\  (thermal: $\log A=8$, $\delta$=0)}    \\
\noalign{\smallskip}
\hline
\noalign{\smallskip}
$\kts$(keV) & 0.28$^{+ 0.02}_{- 0.03}$ & 0.29$^{+  0.02}_{-  0.02}$ & 0.31$^{+  0.01}_{-  0.01}$ & 0.30$^{+  0.01}_{-  0.01}$ & 0.31$^{+0.01}_{- 0.01}$\\
$\kte$(keV) & 2.72$^{+ 0.06}_{- 0.08}$ & 2.65$^{+  0.08}_{-  0.10}$ &2.41$^{+  0.04}_{-  0.04}$  &  2.23$^{+  0.06}_{-  0.05}$   & 2.51$^{+ 0.06}_{-  0.08}$ \\

$\alpha$ & 0.80$^{+ 0.03}_{- 0.03}$ & 0.82$^{+  0.07}_{-  0.03}$  & 0.80$^{+  0.03}_{-  0.03}$ & 0.82$^{+  0.04}_{-  0.04}$& 1.00$^{+  0.08}_{-  0.02}$ \\

$\tau^{\rm b}$ &   6.5 &  6.4 &  6.9 &  7.1  &  5.8 \\
CAF$^{\rm c}$	&   2.8  & 2.6  & 2.5  & 2.4  &  2.2 \\

\noalign{\smallskip}
\hline
\noalign{\smallskip}
\multicolumn{6}{c}{\comptb\ }   \\  
\noalign{\smallskip}
\hline
\noalign{\smallskip}
$\kts$(keV) &  1.00$^{+  0.08}_{-  0.11}$ & 1.15$^{+  0.05}_{-  0.14}$ & 1.11$^{+  0.02}_{-  0.02}$  &  1.08$^{+  0.03}_{-  0.03}$& 1.16$^{+  0.03}_{-  0.04}$ \\

$\rbb^d$(km) & 3.4  & 3.2 & 4.6  & 4.9  & 4.6\\

$\kte$(keV) &  [$\kte$] & [$\kte$]& [$\kte$]  & [$\kte$]   & -\\

log(A)   &  0.11 ($>$-0.12)& -0.40 ($>$-0.65) & [-0.4] & [-0.4] &  [-8]  \\
$\alpha$ & [2.5] & [2.5] &  [2.5] & [2.5] & - \\
$\delta$ &  75 ($>$40) & 49$^{+ 39}_{- 18}$   & 25$^{+  5}_{-  4}$ & 22$^{+  5}_{- 4}$  & - \\
\noalign{\smallskip}
\hline
\noalign{\smallskip}
\multicolumn{6}{c}{\gaussian\ }   \\  
\noalign{\smallskip}
\hline
\noalign{\smallskip}

$E_{l}$(keV) & 	 6.76$^{+ 0.09}_{- 0.08}$ &  6.74 $^{+ 0.11}_{- 0.12}$   & 6.74$^{+  0.09}_{-  0.09}$  & 6.63$^{+  0.13}_{-  0.11}$ & 6.72$^{+  0.10}_{-  0.08}$ \\

$\sigma_{l}$(keV) & 0.3$^{+ 0.2}_{- 0.1}$ &	0.16($<$ 0.33) & 0.10$^{+  0.17}_{-  0.10}$ & 0.12$^{+  0.19}_{-  0.12}$  & 0.06 ($<$ 0.25) \\

$I_{l}$ &	2.5$^{+ 0.9}_{- 0.7}$ & 1.38$^{+  0.73}_{-  0.55}$  &   1.94$^{+  0.71}_{-  0.65}$   &  1.83$^{+  0.95}_{-  0.78}$   & 3.00$^{+  1.73}_{-  0.91}$	\\

$EW_{l}$(eV) & 38$^{+ 18}_{- 9}$ & 17$^{+ 8}_{-8}$ & 15$^{+5}_{-5}$ & 14$^{+ 7}_{-6}$  &  29$^{+ 16}_{-9}$	\\

\noalign{\smallskip}
\hline
\noalign{\smallskip}
\noalign{\smallskip}	
$L_{\rm bb}^e/L_{\rm tot} $ &0.25 & 0.26 & 0.25 & 0.25 & 0.28\\
\noalign{\smallskip}
\hline
\noalign{\smallskip}
$L_{\rm th}^f/L_{\rm tot} $ &0.69 & 0.67 & 0.63 & 0.61 & 0.63\\
\noalign{\smallskip}
$L_{tot}^{\rm g} $ &0.9 & 1.1 &1.7 & 1.7 & 1.5\\
\noalign{\smallskip}
\hline
\noalign{\smallskip}
${\chi^{2}}$/dof & 175/152 & 162/154  &   159/148      & 185/144  & 197/143    \\
\hline
\noalign{\smallskip}

\multicolumn{6}{l}{$^{\rm a}$ In units of 10$^{22}$ cm$^{-2}$.} \\
\multicolumn{6}{l}{$^{\rm b}$ Computed from  $\kte$ and $\alpha$ using equations [17] and [24] for slab geometry in Titarchuk \& Lyubarskij (1995).} \\
\multicolumn{6}{l}{$^{\rm c}$ Compton Amplification Factor (see text for the definition).} \\
\multicolumn{6}{l}{$^{\rm d}$ Computed using the simple relation $L=4\pi \sigma R^2 T^4$.} \\
\multicolumn{6}{l}{$^{\rm e}$ Ratio between the direct BB component flux of the second \comptb\ and the total flux in the 0.1-200 keV energy range. } \\
\multicolumn{6}{l}{$^{\rm f}$ Ratio between the TC component flux and the total flux in the 0.1-200 keV energy range.} \\\multicolumn{6}{l}{$^{\rm g}$ In units of 10$^{38}$ erg s$^{-1}$ computed in the energy range 0.1--200\,keV assuming a source distance of 8\,kpc.}\\
\noalign{\vskip -0.cm}
\label{tab_fit}
\end{tabular}
\end{center}
\end{table*}

\section{Results}
\label{data_fit}
The overall continuum of the five spectra was studied using \comptb\footnote{The \comptb\ model is freely available at web site http://heasarc.gsfc.nasa.gov/docs/xanadu/xspec/newmodels.html}, the newly 
developed Comptonization model described in detail in F08. 
Note that while F08 selected different sources in a given spectral state, 
in this Paper we select a single source and studied its spectral evolution, spanning from the
\intermediate\ state up to the \soft\ state (following the terminology of P06), in order to compare the model parameter evolution  during the spectral transition.  We remind the reader that  Z sources,  unlike atolls, have never been detected in the \lowhard\ or \hardpl\ 
state, possibly  because they never reach low enough accretion rate levels.
Thus the \intermediate\ state and the spectral transition  to the  \soft\ state  is a complete pattern of the  spectral evolution for a Z source.


We briefly remind the reader about the main characteristics of the \comptb\ model \citep{farinelli08}: 
the total emerging spectrum is given by

\begin{equation}
F(E)=\frac{C_N}{1+A}(BB+A\times BB\ast G),
\label{comptb_spectrum}
\end{equation}

where the first and second terms of the right-hand side  represent the seed BB-like photon spectrum
and its convolution with the system Green's function (Comptonized spectrum), respectively.
The factor  $1/(1+A)$ is  the fraction of the seed photon radiation 
directly seen by the Earth observer, whereas the factor $A/(1+A)$ is the fraction of 
the seed photon radiation up-scattered by the Compton cloud. Free parameters of the model 
are the BB seed photons color temperature, kT$_{s}$ and normalization $C_N$,  the plasma temperature, kT$_{e}$,
the logarithm of the illuminating factor A, $\log(A)$.
Moreover, the information on the efficiency of the Comptonization is given by two parameters, 
 $\alpha$ and $\delta$. The parameter $\alpha$ (photon index $\Gamma=\alpha+1$) indicates an 
overall Comptonization efficiency related to an observable quantity in the photon spectrum of the data.
The lower the $\alpha$ parameter (spectrum extending to higher energies) 
the higher the efficiency, i.e. the higher  energy transfer from hot electrons 
to soft seed photons. The $\delta$-parameter provides information about the efficiency of bulk BC with respect that to 
 TC. Both $\alpha$ and $\delta$ are closely 
related (see Fig.~1 in F08) and only in the case of a clear cut-off in the hard X-ray
tail data is it possible to have a precise estimate of the efficiency of both effects.\\ 

\begin{figure}[!th]
\includegraphics[width=9cm, height=12cm]{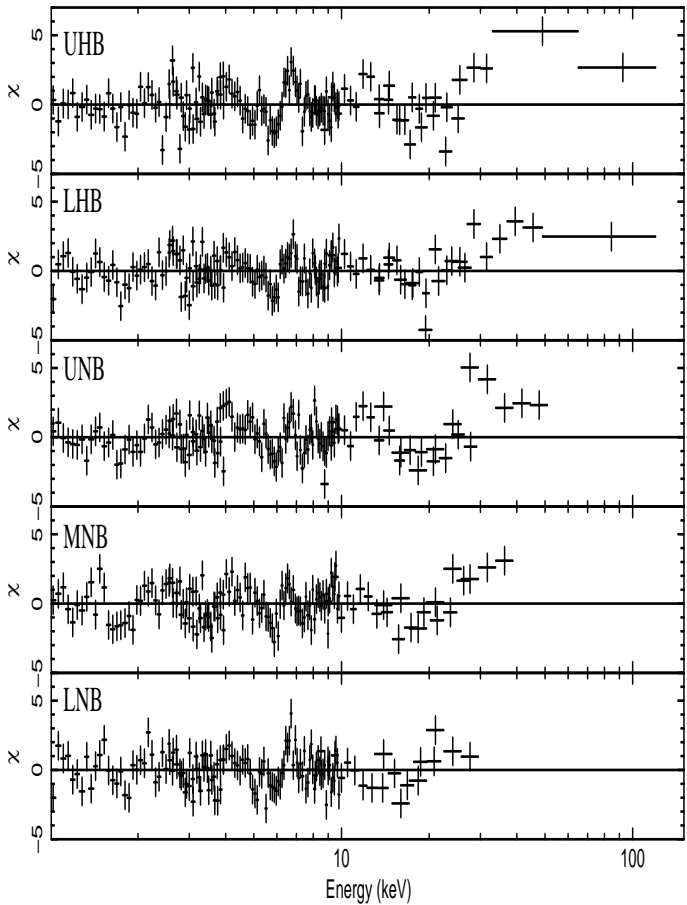}
\caption{Residuals in units of $\sigma$ between the data and best-fit model consisting of a pure TC spectrum (\comptb\ with A$\gg$ 1) and a simple BB (the latter obtained as \comptb\ with $A\ll$1 and 
$\delta=0$). The Gaussian emission lines at $\sim$ 1  and $\sim$ 2.5 keV are also included. 
 The high-energy part ($>$ 20 keV) of the spectrum can be described by simple TC component only at LNB ({\it lower panel}) but the fit  gets progressively worse up towards UHB where systematic deviation of the data is clearly visible.}
\label{residuals}
\end{figure}

\begin{figure}[]
\includegraphics[width=5cm, height=8cm, angle=-90]{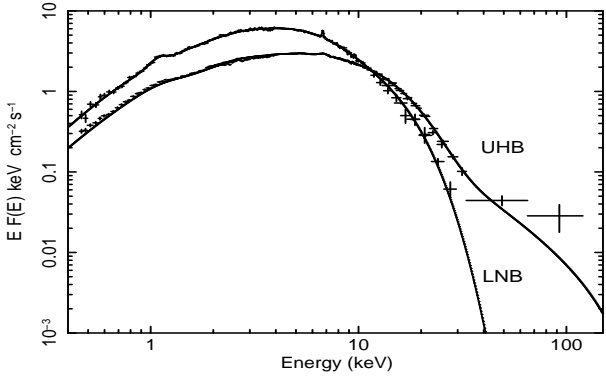}
\caption{Unabsorbed EF(E) spectra and best-fit models (see Tab. \ref{tab_fit}) of the UHB and LNB spectra of \cygx2, which show the spectral evolution of the source from its intermediate state (UHB) to the high-soft state (LNB).} 
\label{spectral_ev}
\end{figure}


Each continuum of the  five spectra of \cygx2\ (two HB and three NB) has been fitted using 
the sum of two \comptb\ components. In Tab. \ref{tab_fit} we report the results of our spectral analysis.
In all cases, the first \comptb\ component [pure TC ($\delta$=0)] provides a significant contribution in  the total emerging spectrum. Most
of the source energetic budget (60-70 \%) is, in fact,  determined  by this component 
which is  described 
(a well-known feature of Z-class LMXBs) by Comptonization of $\sim$ 0.3 keV photons off cool electrons ($\kte \sim$ 3-5 keV) of   high optical depth environment ($\tau \sim$ 10 or 5 depending on the geometry of the cloud, sphere or slab, respectively). The direct seed BB-like spectrum is  not seen in this case because of the high optical  thickness of the Compton cloud.
We note that the BB temperature at UHB stage is a factor about two higher than the value reported in F08 for the same data set. This is due to the fact that here we included a Gaussian emission line at 1 keV which, as already stated in Section 2, may affect both the estimated interstellar absorption $\nh$ and the parameters of low-energy continuum features.\\
In addition to the dominating TC component, broad-band spectra of high-luminosity LMXBs also show the presence of a $\sim$ 1 keV BB-like component whose origin is generally claimed to be close to the NS surface.
This low-energy component is included in our spectra with a second \comptb\ model which, in fact, takes into account  the presence of a directly visible BB-like spectrum (see Eq. \ref{comptb_spectrum}).
In all five spectra there is also an evidence of a Gaussian emission line around 6.7\,keV.
We note however that this continuum model  (TC plus BB) provides accurate description of the data only for the spectrum at  LNB stage, while a systematic excess above $\ga$ 20 keV is observed mainly at UHB and LHB with decreasing  strength of this excess at UNB and MNB (see Fig.~\ref{residuals}). This systematic deviation of  data from the typical soft state continuum model is, in fact, 
the well-known high-energy transient behaviour of some LMXBs, including \cygx2\ (see also DS02).

In general this high-energy feature can be fitted with a simple PL, but in the context of the more physical approach to the X-ray spectral formation  one can treat it  as a BC  emission (see the second term in Eq. [\ref{comptb_spectrum}], BB $\ast$ G, of \comptb). The latter 
describes  the 
result of Comptonization of
a seed photon population by a medium subjected to inward bulk motion so that both TC and BC contribute to the emergent spectrum.
 In this case it is necessary to introduce additional free parameters aimed to describe the physical properties of
the bulk region, namely $\alpha$, $\delta$, $\kte$ and $\log(A)$. 

However, if the number of free 
parameters describing the high energy part of the X-ray spectrum (i.e. the second \comptb\ in Tab. \ref{tab_fit})
exceeds the number of observable quantities (PL slope, normalization and spectral
cut-off), degeneracy of the parameters (impossibility to constrain them) occurs and some assumptions have to be made in order to reduce the free-parameter space dimension.
As the first of our assumptions  we suggest that 
the Comptonizing plasma temperature of the two \comptb\ models are equal each other (see Tab. \ref{tab_fit}). In fact  as already pointed out by  F08 there is no expectation of some strong electron temperature  variation in the region extending from the outer TC medium up to the innermost bulk region.
The lack of an unambiguous rollover ($\ga $ 100 keV) in the high-energy data reduces the number of measurable spectral characteristics  of the high-energy transient X-ray emission, which are essentially the PL slope  and its normalization. The latter quantity is expressed through $\log(A)$ in the model, while the spectral index  $\alpha$  
provides the spectral slope. For a given fixed value of $\alpha$, 
when  the bulk parameter $\delta$ increases, the high-energy cut-off energy increases too, leading to the extension of the PL-like component in the spectrum (see Fig.~1 in F08).
 
The statistics of the spectral  data  at UHB and LHB stages did not allow us to simultaneously constrain $\alpha$, $\log(A)$ and $\delta$. Since  we are mainly interested in studying the importance of BC  as a function of the source position in the Z-track, this mapping was possible only leaving free the bulk parameter $\delta$ and fixing the energy index $\alpha$. A detailed study of the parameter space showed us that the minimum of the \chiq\ could be obtained by fixing  $\alpha$=2.5. The results obtained with this modeling
 approach  are reported in Tab. \ref{tab_fit}.  One can see  that only the lower limit (at 90\%
 confidence level) on $\log(A)$  can be determined  in  both HB spectra, while as far as $\delta$ is concerned, we have a lower limit for its
value at UHB and better constrain of this  at LHB.
The situation gets even more critical for the parameter determination  of the NB spectra.
 In the UNB and MNB cases, where a high-energy excess is still observed over the BB plus TC continuum (see Fig.~\ref{residuals}) it was not possible to put serious constraints on both $\delta$ and 
 $\log(A)$ leaving them simultaneously free, mainly because of the lack  of counts at high energies. 
The only possibility  to explore the  behaviour of $\delta$ was thus fixing  $\log(A)$ at the 
best-fit value of the LHB spectrum (see Tab. \ref{tab_fit}). Finally, in the LNB spectrum 
the second \comptb\ model requires neither thermal nor bulk Comptonization, resulting in  
a simple BB shape. The overall LNB spectrum could  hence be well fit by the sum of thermal Comptonization 
(the first \comptb) plus a BB.
In terms of best-fit improvement, the addition of the TC component to the  BC one  gives $\Delta \chi^2 \sim 100$ (F-test $\sim 10^{-17}$) and  $\Delta  \chi^2 \sim 63$  (F-test $\sim 10^{-10}$) for the UNB and MNB spectra, respectively, while for the LNB spectrum it is  only $\Delta  \chi^2 \sim 19$ (F-test $\sim 10^{-4}$).

\begin{figure*}[!th]
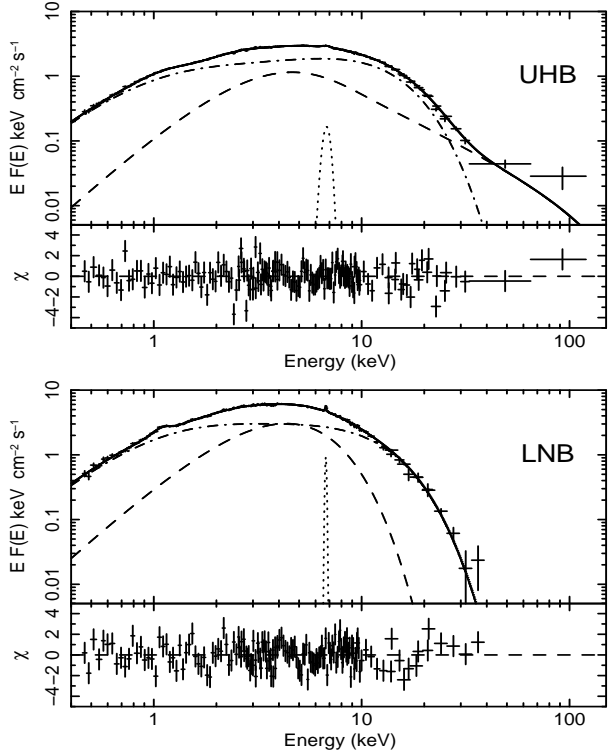

\begin{center}
\includegraphics[width=5cm, height=8cm, angle=-90]{fig6.eps}
\includegraphics[width=5cm, height=8cm, angle=-90]{fig7.eps}
\caption{Unabsorbed EF(E) spectra, best-fit model  and residuals between data and model in units of $\sigma$ for the UHB and LNB spectra of \cygx2. Different line styles
represent single component of the best-fit model reported in Tab. \ref{tab_fit}. {\it Dotted-dashed}: \comptb\ with pure TC. {\it Long-dashed}: \comptb\ which includes TC and BC, i.e. $\delta\gg1$ ({\it  left panel}) or with simple BB spectrum ({\it right panel}). {\it Dotted}:Gaussian emission line.} 
\label{efe1}
\end{center}
\end{figure*}

 \section{Discussion}
We have studied newly extracted \cygx2\ \sax\ spectra, covering the horizontal and 
normal branch of the Z-track, using the \comptb\ model 
recently developed by \cite{farinelli08}. 
We remark that \comptb\ is a Comptonization model that includes both TC and BC, namely it can be considered as an (updated) merging of the well known \comptt\ and \bmc\ \xspec\ models. \\

\subsection{Interpretation of the observed spectra}
We have fitted the continuum of all the five source spectra  with two \comptb\ models.
The first one represents the dominating TC component of the spectrum,
while the second one provides description of the strong $\sim$ 1 keV  BB-like feature and, when observed,
the hard X-ray emission above 30 keV.
 In our scenario (see also F07 and F08), the 
thermal \comptb\ component  is an emission originated in the relatively cold $(\sim$3\, keV)  and optically thick  
($\tau\sim$ 5) outer transition layer due to the Comptonization of cold disk seed photons 
($\sim$ 0.3\,keV). This component is rather stable as can be seen in Tab. \ref{tab_fit}.
The second \comptb\ component originates in the innermost region of the system and is subjected to major changes (Tab. \ref{tab_fit}). 
Hot BB photons ($\sim$1\,keV) are emitted by the neutron star surface and within the transition layer 
itself. The spectral shape variability is due to the  changing Comptonization 
of this seed photon population, mainly led by the accretion rate, as discussed Section~\ref{evolution}. 
The difference in the measured values of $\kts$ (a factor about three) strongly points in favour of two distinct populations 
of seed photons, which in turn translates in the existence of two physically separated 
regions.
Looking at the best-fit parameters in Tab. \ref{tab_fit}, the source spectral stability of the persistent continuum is almost evident, in general agreement with high-luminosity LMXBs properties.
Moreover, in the MNB and LNB spectra, the TC parameter are also well constrained, differently from what reported in DS02 where they strongly deviate from a smooth trend with unreasonable associated errors.
We claim that one reason of the DS02 results could be the rough timing selection for spectral analysis
(see Section 2).
The quantity that undergoes the major change as a function of the source position in the CD/HID is in fact the estimated  
0.1--200 keV luminosity which has a maximum increase of about 80\%  from the UHB to the UNB-MNB state.
Less significant variations are observed in the plasma temperature $\kte$ and the thermal component \emph {Comptonization amplification factor} (CAF),  defined as the ratio of energy fluxes of Comptonized over seed photons. 
The CAF, in fact, decreases monotonically from the UHB to the LNB. We also observe that the inferred  plasma temperature, optical depth and
luminosity in the LNB slightly deviates from the monotonic trend seen in the first four spectra. Whether this small jump is real or due to
some bias in the spectral modeling is not straightforward. In particular the presence of a strong 1 keV and weaker 2.5 keV  emission feature
especially in the NB spectra may play some role in the actual determination of the temperature of the seed photons of the pure thermal
\comptb\ model, which is reflected in the $\kte$ and $\alpha$ values.  \\


\subsection{Spectral evolution as a function of mass accretion rate}
\label{evolution}

The new theoretical and observational support provided in the last years (e.g., TMK97, P06, F08) to the 
study of hard X-ray emission in LMXBs significantly pushed forward our knowledge about the nature of 
the emission processes in these sources.
Multi-wavelength observations of Z sources \citep[e.g.,][]{penninx88, vrtilek90} strengthened the evidence that the  motion of the sources along the Z track is mainly controlled by the accretion rate $\dot{M}$ which   increases from the HB to the FB.
However, little is said on why an increase in $\dot{M}$ leads to a disappearance of the hard tail in LMXBs 
and, moreover, the actual
meaning of $\dot{M}$ itself is not fully understood. It is however widely accepted that the simple source bolometric X-ray luminosity is not a good $\dot{M}$-tracer, 
a hint occurring not only in NS but also in black hole (BH) systems \citep[e.g.,][]{vdk01}.
We propose to actually define \emph{two} $\dot{M}$-values for a single source, one related to the accretion disk
$\dot{M}_{\rm disk}$ and another one ($\dot{M}_{\rm tl}$) to the so-called transition layer (TL), the region where the disk angular velocity deviates from its Keplerian rotation in order to adjust to the angular velocity of the slowly spinning NS.
The reason for this splitting is that at very high accretion rate levels (close to the Eddington limit), the radiation pressure from the accretion disk may eject a significant fraction of the accreted matter, producing a 
powerful wind surrounding the system \citep{bradshaw07}. In this case the mass flow coming from the innermost part of the  disk through the TL to the NS surface is less than that arriving at the disk outer part, namely  $\dot{M}_{\rm tl} \la \dot{M}_{\rm disk}$, with $\dot{M}_{\rm tl} \propto \dot{M}_{\rm disk}$ \citep{tsa07}. 	 

Focusing the attention to the TL region, it is worth noting that both theoretical  (TMK96, TMK97) and extended data analysis works (P06, F07, F08) led to the possibility that the bulk Comptonization process in the TL is responsible for the  hard X-ray emission in NS LMXBs. 
A self-consistent  physical treatment of the TL region is presently being developed by  Titarchuk \& Farinelli (2008, in preparation, hereafter TF08); the numerical solution of the radial momentum equation performed by FT08 unambiguously shows that matter arriving at the \emph{transition radius} from the Keplerian disk  with some fraction of the characteristic disk radial  (magneto-acoustic) velocity and then it proceeds through the transition layer  towards the NS  with an almost constant velocity up to some 
radius over which it further proceeds  in quasi free-fall manner, getting the NS surface with   $ V_{R} \sim 0.1$c.   
However, as FT08 emphasize,  this free-fall region \emph{is not always present} but it depends on \emph{ the mass accretion rate $\dot{M}$}, which can thus be now identified with  $\dot{M}_{\rm tl}$.
In particular, high $\dot{M}_{\rm tl}$ generates a strong radiation field in the TL 
whose pressure gradient prevents  matter from reaching the NS surface with the 
high-speed which is necessary to produce, via Doppler effect, hard X-ray photons (hence no hard tail detected).
We remind once again the reader that the only observable quantities in the high-energy  X-ray spectra 
are the  PL photon index $\gamma=\alpha+1$,  its normalization and the high-energy cut-off, if detected.
The actual accretion rate (regardless its definition) is not a directly measurable 
quantity; its estimation can be done only including it as a free parameter in devoted accretion models. This is, e.g., what has been done by \cite{bw07} in their model for accretion columns in X-ray pulsars.
On the other hand, in \comptb\ the accretion rate is not a free parameter, it is just somewhat implicitly hosted in the bulk parameter as $\delta \propto \dot{m}^{-1}$ (where $\dot{m} \equiv \dot{M}/\dot{M}_{\rm Edd}$, F08).
We also  note  that in the \emph{intermediate state} spectra of NS LMXBs,  \emph{two spectral indexes} are actually observed in the X-ray spectrum, the first one related to pure TC and the second one to a mixed thermal-bulk Comptonization effect.


For \mbox{Cyg X--2}, in the former case  the spectral index $\alpha \sim$ 1 (see Tab. \ref{tab_fit}) which is  typical  for the intermediate state in  NS sources \footnote{This result can be verified from the best-fit values of $\kte$ and $\tau$ obtained from \comptt\ in previous works 
(see references in Section 1) and using Eqs. [17] and [24] in TL95.}.
As far as the second spectral index is concerned, it is important to discuss the differences between 
systems hosting 
a NS or a BH.
For BH sources, subjected to enhanced spectral transitions, in fact just {\it one} index is measured in the 
spectrum. It evolves from $\alpha \sim 1.7$ in hard state, which is TC-dominated with observed cut-off around 
100 keV, to  $\alpha \sim 2.8$ in the BC-dominated soft state. In this BC case, {\it index saturation} is 
observed (Shaposhnikov \& Titarchuk 2008). In fact, it can be  proven \citep{bradshaw07} that 
$\alpha \approx Y^{-1}$, where the Comptonization parameter $Y = \eta N_{\rm sc}$, $\eta$ is the average  energy exchange per scattering and $N_{\rm sc}$ is the average number of scatterings. In a BC-dominated 
flow, $\eta \propto \tau^{-1}$ and $N_{\rm sc} \propto \tau$ \citep{lt07} so that 
saturation occurs. But this actually occurs in BH systems, where  the presence of a fully absorption inner 
boundary condition  for the radiation  (TMK97) allows bulk motion to be always present.
In the case of NS system, the presence of a solid surface (reflection boundary condition) plays a 
crucial role, given that the radiation pressure gradient gets increasing with the accretion rate until 
bulk flow is stopped.
Thus no index saturation can be observed  in NS sources. The innermost part of the system evolves towards thermal  equilibrium when the emergent spectrum consists of two blackbody-like components which  are  related to NS surface and disk  emissions \citep{ts05}.

On the basis of these considerations we may tentatively discuss the parameters behaviour reported in 
Tab. \ref{tab_fit}: along  the horizontal branch (HB), at lower
$\dot{M}$ level, BC effect can efficiently pronounced as it is indicated  by the high $\delta$-value. A slight decrease of the BC effect is observed in the  spectra in the low part of HB (LHB), where both $\delta$ and $\log(A)$ get lower, even though this effect seems to be marginal given the  high error bars  for both these parameters. The total (0.1-200 keV) source luminosity at LHB increases about 20\% with respect to that at UHB. When the sources is at the UNB (upper normal branch), a further increase ($\sim$ 50 \%) of the luminosity with respect to that at LHB occurs, accompanied by significant decrease of the hard X-ray emission (see Fig.~\ref{residuals}).
We may interpret this behaviour as an increase of $\dot{M}_{\rm disk}$ which in turn leads to higher $\dot{M}_{\rm tl}$ and consequently leads  to the increase  of the radiation pressure in the TL. 
This pressure works against an inward matter motion and the bulk effect decreases (though it may be still present but below the instrument threshold). Figure \ref{spectral_ev} illustrates this spectral transition, due to  $\dot{M}$ change,  from the intermediate (UHB) to soft (NB) state, following the terminology defined in P06.

In the spectrum the power-law  extension  reduces  as a result of the BC efficiency  ($\delta-$parameter) reduction. Looking at the parameters of Tab. \ref{tab_fit} and at the residuals in Fig. \ref{residuals} it is possible to see that the situation only slightly changes at MNB. 
Finally, in the LNB  no hard X-ray emission is observed and a simple BB-like spectrum along with  the TC bump is present. It is possible that at this stage the effect of radiation pressure in the TL is strong enough in stopping bulk motion, or at least bringing its visible effect below the instrument threshold. At this stage it is very likely that matter follows a very complex behaviour; it may arrive at the NS surface very slowly or it may significantly spread over the stellar surface \citep{is99} producing additionally a thick layer which suppresses the radio emission (P06). 

Note however that the 0.1-200 keV luminosity at  LNB  is about 12\% lower than that at UNB and MNB.
These results clearly demonstrates that  the X-ray luminosity is not a good tracer of the accretion rate, whatever one defines it. 
In all five spectra, a direct BB-like component, providing about 25\% of the source energetic budget, 
is observed (indeed in the second \comptb\ component, $\log(A)$ is small in all cases). The estimated BB \emph{color radius} in all cases is $R_{\rm bb} \la$ 5 km; while this value has to be treated very carefully (it is derived from the measured BB \emph{color temperature}  $\kts$ assuming isotropic emission) its order of magnitude is consistent with the NS radius, pointing to an origin close to the compact object surface.
In our interpretation, these BB-like seed photon spectra have origins related to the thermal emissions of  the NS surface and to local energy release emission  in the TL region. It is possible  that part of this soft component is direct radiation from the seed photon region (NS or TL) and the other one  consists of photons which  are not affected by  the  Comptonization.
The fraction of this directly escaping radiation is parameterized through $\log(A)$ related to both a geometrical  (covering factor) and spatial (seed photon distribution) system configuration in the innermost part, even though the relative importance among the two effects cannot  be determined from the X-ray spectrum.

\section{Conclusions}

The scenario where both thermal  and bulk Comptonization contribute to 
explain the  LMXB spectral evolution 
is gaining strong theoretical
\citep[TMK96, TMK97,][TF08]{lt99,lt01}  and 
observational support \citep[][P06, F07, F08, this paper]{shrader98,borozdin99}. 
This is true also for the case of accretion  powered \mbox{X-ray} pulsars for which a 
new theoretical model based on  
thermal and bulk Comptonization occurring in the accreting shocked gas  
has been recently presented \citep{bw07}.
In the model presented by \cite{bw07}, the effects of the strong magnetic field 
($B \sim 10^{12}$\,G) become very important and are included, whereas 
 in this paper we have used a model that has been recently proposed 
by F08 to study low-magnetized compact objects (B$\la 10^9$ G), where effects 
of the magnetic field within the flow can be neglected (as opposed to the radiation pressure).
The two models are complementary and we believe they can be an important step forward
to understanding and relating the 
physics of bursting (low-$B$) and pulsating (high-$B$) accreting NSs.
Nevertheless the present analysis has revealed the key importance of having a next generation of very high-sensitivity 
missions at energies extending up to 300-400 keV. 
The detection of high-energy cut-off in extended PL-like tails of both NS and BH binary systems would be one breakthrough in understanding the physics of accretion processes very close to the compact object, providing
constraints on the bulk energy of the matter.

\begin{acknowledgements}

The authors are very grateful to the anonymous referee, whose suggestions strongly improved the
quality of the paper with respect to the formerly submitted version.
AP acknowledges the Italian Space Agency financial and 
programmatic support via contract I/008/07/0. 
This work has been partially supported by the grant from Italian PRIN-INAF 2007, "Bulk motion 
Comptonization models in X-ray Binaries: from phenomenology to physics", PI M. Cocchi.

\end{acknowledgements}

\bibliographystyle{aa}
\bibliography{biblio}

\end{document}